\documentclass[conference]{IEEEtran}
\IEEEoverridecommandlockouts
\usepackage{cite}
\usepackage{amsmath,amssymb,amsfonts}
\usepackage{algorithmic}
\usepackage{graphicx}
\usepackage{textcomp}
\usepackage{xcolor}
\usepackage{cite}
\usepackage{subfigure}
\def\BibTeX{{\rm B\kern-.05em{\sc i\kern-.025em b}\kern-.08em
    T\kern-.1667em\lower.7ex\hbox{E}\kern-.125emX}}
\begin{document}

\title{Infnote: A Decentralized Information Sharing Platform Based on Blockchain}

\author{\IEEEauthorblockN{Haoqian Zhang$^1$, Yancheng Zhao$^2$, Abhishek Paryani$^3$, Ke Yi$^4$}
\IEEEauthorblockA{Computer Since \& Engineering Department\\
\textit{The Hong Kong University of Science and Technology} \\
Hong Kong, China \\
Email: \{$^1$haoqian.zhang,$^3$aparyani\}@connect.ust.hk, $^2$yanchengz@ust.hk, $^4$yike@cse.ust.hk}
}


\maketitle

\begin{abstract}
Internet censorship has been implemented in several countries to prevent citizens from accessing information and to suppress discussion of specific topics. This paper presents {\em Infnote}, a platform that helps eliminate the problem of sharing content in these censorship regimes. Infnote is a decentralized information sharing system based on blockchain and peer-to-peer network, aiming to provide an easy-to-use medium for users to share their thoughts, insights and views freely without worrying about data tampering and data loss. Infnote provides a solution that is able to work on any level of Internet censorship. Infnote uses multi-chains architecture to support various independent applications or different functions in an application.
\end{abstract}


\section{Introduction}
\label{sec-introduction}
Freedom of speech is considered a basic human right under Article 19 of the \textit{Universal Declaration of Human Rights} \cite{assembly1948universal}. The evolution of digital telecommunications have brought with it both opportunities and challenges for freedom of speech. On the one hand, we can access or deliver information faster via several mediums, while on the other hand, regulators can use both technical and non-technical methods to control or suppress what can be published or viewed on the Internet. While Internet users can utilize circumvention technologies to bypass Internet censorship to access or publish information, regulators around the world have significantly increased their efforts to control the information flow on social media \cite{house2017freedom}. A recent report concludes that the internet is becoming more restricted globally, and democracy itself is withering under its influence \cite{house2018freedom}.

The current Internet infrastructure model is heavily centralized. There are only 13 logical DNS root name servers and IP address space is directly controlled by ICANN. These are apex players that control the Internet and are involved in delivering messages to the masses. This is being challenged by complex censorship techniques such as DNS and IP blocking and hacking attacks on content hosting websites (e.g. blogs, social media platforms and more). Hence, there is an urgent need to solve the challenges related to this blockade by employing better circumventing approaches.

As we review in Section \ref{sec:CircumventionTechnologies}, there are many techniques that allow users to circumvent Internet censorship. However they all come with their own pros and cons. Methods based on proxy or VPN are commonly used. However, they rely on the connections to servers hosted in a country with less or no censorship and the connections may fail due to reasons like deep packet inspection or white-listing on gateway firewalls. Sneakernet allows a user to access information under any level of censorship, but it cannot guarantee validated information. Peer-to-peer networks provide an approach to transferring data within the boundaries of censorship regions and there is no single point of failure, making it difficult to block. ZeroNet \cite{Zeronet} is a peer-to-peer web hosting project, based on the BitTorrent protocol, which can be used to circumvent Internet censorship, but the ZeroNet site owners have full control over the content of their websites. This paper will apply blockchain techniques to circumvent Internet censorship. 

Bitcoin caught everyone's attention when it appeared in a white paper in 2008 \cite{nakamoto2008bitcoin}; various cryptocurrencies based on blockchain have emerged since then, with some improving bitcoin and others being more innovative, like Ethereum and Hyperledger. Today, the applications of blockchain and their respective peer-to-peer (P2P) networks are designed to function for much more than a decentralized currency. 

Blockchain, as an append-only global ledger, has already been used in a decentralized version of the DNS\cite{loibl2014namecoin} and data storage. The append-only ledger is ideal for an information sharing platform aimed to circumvent Internet censorship, since no one would have the authority to delete the content stored in the ledger. IPFS \cite{benet2014ipfs} and Blockstack \cite{ali2016blockstack} are two existing data storage platforms based on blockchain. However, IPFS currently does not support the publish-subscribe pattern, and Blockstack utilizes centralized cloud servers to store data \cite{ali2017blockstack}. In China, people have already started to use blockchain to battle government censorship. For example, in 2018, an anonymous user attached an open letter to an ether transaction and posted it to the Ethereum blockchain \cite{MetooBlockchain}. 

Infnote, is our answer to the limitations of the existing platforms. The name 'Infnote' comes from providing \textbf{inf}inite power through the \textbf{note}s that the community of users publish. Infnote provides a tool for content creators, social activists, journalists and others who simply want their voices to be heard. 

Infnote, based on blockchain and P2P technologies, aims at providing a platform for users to share their thoughts, insights and opinions under varying levels of Internet censorship. It is a decentralized platform that can provide the user with pseudo-anonymity and transparency, and allow content to travel and be viewed freely across a network of users. Unlike conventional blockchain, that uses a single chain to store information, Infnote uses multi-chains to support various independent applications or different functions in an application.

In this paper, we first introduce the current situation of Internet censorship around the world and define its different levels, mainly from a technical point of view in Section \ref{sec:InternetCensorship}. Different circumvention technologies are then compared in Section \ref{sec:CircumventionTechnologies}. Next, several related projects are introduced in Section \ref{sec:RelatedProjects}. Then, various popular consensus mechanisms are analyzed in Section \ref{sec:Blockchain}. The detailed design and implementation of Infnote is presented in Section \ref{sec:Designandimplementation}. The performance and the results of evaluation are demonstrated in Section \ref{sec:Evaluation}.

\section{Internet Censorship}
\label{sec:InternetCensorship}

The Internet is supposed to provide an open platform that allows anyone to share information, access opportunities and collaborate across geographical boundaries\cite{solon2017tim}. However, the Internet is being challenged today by political systems around the world. Information flow is being manipulated to show propaganda, while sources of real information are either blocked or redacted, and in many cases citizens remain unaware of the happenings outside their borders. On the one hand, we have the concept of an open, decentralized, democratized Internet and on the other hand we also have the Great Firewall of China, the Halal Internet from Iran, the Kwangmyong intranet from North Korea. 

It is this type of \textbf{censorship} for this paper hopes to address. Before presenting our solution, we first briefly review some of the censorship methods commonly deployed by various countries.

\subsection{Censorship Methods}

\paragraph{DNS Manipulation or Tampering}
In oppressive countries, if the regulator wants to censor a website, they can employ a technique called DNS manipulation or tampering: when a client requests an IP address, the DNS server sends back a false IP address, meaning the client is actually visiting an incorrect website.

\paragraph{Domain and IP Address Blocking}
One of the methods to block a user from accessing a website is to block its domain and IP at the Internet gateway level.

\paragraph{Throttling}
An ISP can control the traffic and speed, which is known as bandwidth throttling. In some countries, this technique is used during political events\cite{anderson2013dimming, aryan2013internet}. From a technical perspective, throttling is achieved by slowing down TCP either by dropping packets \cite{fifield2017threat} or controlling the bandwidth provided to a specific protocol.

\paragraph{Deep Packet Inspection (DPI)}
Deep packet inspection is another form of packet filtering that is used heavily in certain countries for the purposes of monitoring, blocking and sometimes throttling data flow through the Internet gateway systems. DPI filtering is used by Internet service providers to scan the payload of the Internet packets along with a normal scan of the headers to determine how to classify and control it, and whether or not to drop it. This is possible in real-time with the equipment that is available today.

\paragraph{Content and Keyword Filtering}
Politically repressive countries pro-actively block foreign news websites, pornography, propaganda websites and content that do not match their political principles and philosophies. One easy way of censoring websites is based on their content, domain name, and specific keywords. Any website that matches a specific criteria or filter, is automatically censored for violation of the governments policies. 

\paragraph{Distributed Denial of Service (DDoS)}
This type of censorship method has been used in the past to take down websites that have taken a stand against a regime \cite{marczak2015analysis}. From a technical perspective, multiple computers on the network are controlled either deliberately or unwittingly and a coordinated series of traffic is sent to a target server or cluster of servers in the cloud. The traffic could be in the form of either ICMP, UDP packets, SYN flooding or a combination of those types of traffic to exhaust the computer resources of the target.

\subsection{Levels of Internet Censorship}
\label{sec:levelsofinternetcensorship}
We provide a context to the censorship problem by defining the degrees of censorship. In section \ref{sec:Effectiveness}, we will analyze our system in different levels of censorship.

\paragraph{Level 1: Little or No Censorship}
Little or no censorship is enforced in these countries. There is no need to use any circumvention technology.

\paragraph{Level 2: Selective Censorship}
A small number of websites are blocked. Simple censorship methods, like IP address blocking or DNS filtering and redirection are likely to be used. Websites dealing with illegal or illicit activity may be blocked. Citizens can easily use circumvention technology to bypass the censorship. 

\paragraph{Level 3: Substantial Censorship}
A large portion of content on the Internet is blocked and several censorship methods are implemented simultaneously. A blacklist of IP addresses and domains is likely to be enforced by the firewall, filtering Internet traffic that goes through the border Internet gateway systems. Anti-censorship circumvention tools may also be targets of censorship, making it extremely difficult for citizens to bypass the censorship. 

\paragraph{Level 4: Pervasive Censorship}
At this level, a whitelist is enforced by the firewall at regional boundaries, implying that only approved Internet traffic will be allowed to pass the boundary of a censored region. This makes it theoretically impossible to use any proxy or VPN, whose IP is outside of the boundary.

\paragraph{Level 5: No Internet}
In extreme situations, the Internet service may be completely cut off and circumvention tools that rely on the Internet will not work. It is extremely difficult for citizens to access or distribute digital information. During the Arab Spring, the Egyptian government shut down the Internet in Egypt temporarily.

\section{Circumvention Technologies}
\label{sec:CircumventionTechnologies}

Given the censorship scenarios and the evolution of such censorship systems on the web, there has been an uptake in anti-censorship technologies as well. Some circumvention technologies are simple, while others require more advanced knowledge of systems to implement and make them work.

\subsection{Introduction to Circumvention Technologies}

\paragraph{Cached Pages}
Search engines like Google or Archive.org save or cache pages through its set of crawlers. Users can simply search for a web page and access its cached versions. This is an easy and quick way to access blocked and censored content. Archive.org saves multiple versions of a web page so a user can access the past versions, and even a web page that has been taken down or gone offline can be accessed via this service. However, these websites and services can also be blocked by censors, making this circumvention method effective only for Level 1 and Level 2 censorship.

\paragraph{Proxy}
A proxy server is a server that sits in between a client (requesting information, content, images etc.) and a server (that contains the information). A proxy server needs to be configured on the user's browser or application. A proxy can provide encryption and other forms of security to the user. Users can also access a proxy website based on a proxy server that is hosted in a country with less or no censorship. Once you input the URL you want to reach, the proxy websites fetch the content and displays it. However, the regulator can easily ban these proxy servers, and hence this method is only effective for Level 1 and Level 2 censorship.

\paragraph{Virtual Private Network (VPN)}
Initially, VPNs were used to access internal networks (e.g. office intranet) from the public Internet. Recently, we have seen rapid growth in the deployment of VPN's to circumvent Internet Censorship \cite{zhang2004overview}. A VPN works by creating a virtual end to end connection through virtual tunnel protocols. By using a VPN, a user residing in a censorship regime can access blocked content by setting up a secure connection to another country with little or no Internet censorship. However, it requires additional software to set up connections and the software may be banned in substantial censorship regimes.

\paragraph{Peer-to-Peer (P2P) Network}
One widely used P2P network is BitTorrent, which is a robust protocol for file-sharing that allows users to download content from multiple sources in a swarm that contains seeders, peers and leechers. From a technical perspective, BitTorrent breaks down a single file into several pieces or chunks of data. Peers then pull bits of files from seeders and other peers\cite{cohen2008bittorrent}.

Anonymous networks like Onion Router (Tor) \cite{TorProject} and Invisible Internet Project (I2P) \cite{I2PProject} offer peer-to-peer communication through their censorship-resistant and anonymous networks by relaying traffic through multiple nodes. They are often open-source and use circuit based systems that encrypt the user's traffic end-to-end so that neither the sender nor the receiver need to reveal their respective IP addresses. It supports multiple applications like instant messaging, web browsing, file-sharing and so on. However, they relies on centralized servers and the speed is lower than direct connection because of the multiple hops.

\paragraph{Sneakernet}
Sneakernet is to physically transport digital information from one physical location to another, thereby helping distribute information to other users or groups and circumventing surveillance and censorship. This method can work in countries with no Internet or pervasive censorship deployed given its lack of little reliance on the Internet. The major drawback is that the source of information and the content itself cannot be fully trusted, and this is a slow method of communication. 


\paragraph{Web-to-Email}
This is a simple service that takes a snapshot of any website and sends it directly to the users' email. In geographies with little to selective censorship, this method can be quite effective. One disadvantage is that if the emails and the website to access this service are blocked, it will no longer be effective. 

\subsection{Features of Circumvention Technologies}

\begin{table*}[tbp]
\centering  
\begin{tabular}{lcccccc}  
\hline
 & Cached Pages & Proxy & VPN & P2P & Sneakernets & Web-to-Email\\ 
\hline 
\texttt{Difficulty (identify \& block)} & Easy & Medium & Medium & Hard & Hard  & Easy \\
\texttt{Anonymity} & No & No & No & Yes & Yes & No \\
\texttt{Data Tampering Protection} & No & Yes & Yes & Yes & No & No \\
\texttt{Encryption} & No & Yes & Yes & Yes & No & No  \\
\texttt{Censorship (apply to)} & Level 2 & Level 3 & Level 3 & Level 4 & Level 5 & Level 2 \\
\hline
\end{tabular}
\newline
\caption{Censorship Technologies - Comparison Table}
\label{tab:censorship-tech-compare}
\end{table*}

The various circumvention methods are compared to understand them under the following criteria in \textbf{Table \ref{tab:censorship-tech-compare}}.

\paragraph{Difficulty Level - Identify and Block (Rating: Easy, Medium or Hard)}
This rating refers to how easy or difficult it is for the government, regulators, and Internet service providers to block the techniques on the Internet. For example: To block foreign websites that provide news, the government has to block the IP address or the domain name of the website. This is considered an easy task for the government compared to the feature-set of sophisticated firewalls and backbone systems in their inventory.

\paragraph{Anonymity (Rating: Yes or No)}
This rating refers to whether anonymity could be provided to users by the circumvention tool or method. 

\paragraph{Data Tampering Protection (Rating: Yes or No)}
Data like files, html pages, music, videos and so on, can be tampered with once they are away from their original source. Only in some cases, through techniques like hashing or digital signature can one be assured that the data is from the original source and that it has not been changed or modified in any way.

\paragraph{Encryption (Rating: Yes or No)}
To prevent any unauthorized access, users can encrypt data with algorithms. This classification provides a simple yes or no to the following question: can the data be encrypted while being stored at the source and viewed by decrypting it?

\paragraph{Censorship Category (Rating: Level 1 to 5)}
Each circumvention technology has its limitations when it comes to deceiving the censors or simply finding another route to access or distribute content. The category level (1 to 5), as defined in Section \ref{sec:InternetCensorship}, at which the circumvention technology can be effective, is mentioned through this feature. Each level is consecutive in nature and hence includes features of the previous level.

\section{Related Projects: Peer-to-Peer Web Hosting}
\label{sec:RelatedProjects}

Infnote is a peer-to-peer web hosting project, that uses peer-to-peer networks to distribute and access web pages without the need for any intermediary hosting providers. This feature makes this technique very suitable for circumventing Internet censorship. In the following, we review some popular and working systems related to Infnote.

\paragraph{Interplanetary file system (IPFS)}
IPFS is a distributed file storage system \cite{benet2014ipfs} that combines the power of decentralization and the web by providing features like strong data integrity by using merkle trees and distributed hash tables, and availability by replicating data across the networks \cite{vargascensorship}. Hashes of content are stored within the blockchain \cite{ali2017iot}. It is a good platform for evading censorship because access to it is difficult to block\cite{alabdulwahhab2018web}. However, IPFS does not support the publish-subscribe pattern, which is necessary for real-time information sharing (pubsub is an experimental implementation for it). It would have been possible to implement Infnote on top of IPFS. However, in order to gain more flexibility, we decided to design our architecture from the ground up. 

\paragraph{FreeNet}
FreeNet, similarly to IPFS, uses a distributed data storage mechanism where the storage space is distributed amongst all nodes on the network. It is decentralized, thereby making it difficult to take down in censorship-driven countries \cite{clarke2001freenet}. However, unpopular files might disappear from the network \cite{hasan2005survey}, which does not fit the use case of information sharing platforms for which browsing old posts is needed. With Infnote's architecture, where we use blockchain, we can guarantee that data cannot be removed or lost from the system.

\paragraph{ZeroNet}
ZeroNet is based on a file system and BitTorrent protocol where sites are recognized through a public key,  unlike the traditional web where sites are recognized by domains and IP addresses \cite{Zeronet}. The site owner can sign files with their private key to make modifications and authorize other users to make modifications to support multi-user sites. Each user requests other nodes' addresses from the BitTorrent tracker and shares or receives the site files with them. As long as a site is supported by peers, the content is alive and can be accessed by ZeroNet users. However, a site owner has full control over the content of the website, resulting in excessive power residing with the site owner. Currently, sites cannot be directly accessed from web browsers unless by using the ZeroNet application, and the history of modification is not stored. Also due to the ZeroNet BitTorrent trackers being blocked, ZeroNet sites are inaccessible in China. In Infnote's case, the site owner's power is limited as no one can delete the data in the blockchain and users can directly access content from their respective web browsers. In addition, Infnote can work under different levels of censorship and support various approaches to find and obtain content from peers, making the system difficult to block.

\paragraph{Blockstack}
Blockstack is a project that provides decentralized dictionary storage, similar to Namecoin \cite{loibl2014namecoin}, built on top of the bitcoin blockchain. It is a strong solution for deploying a decentralized public key infrastructure (PKI), as demonstrated in \cite{ali2016blockstack}. However, one disadvantage is that the values are stored in a centralized cloud storage system \cite{ali2017blockstack}, which makes it less ideal for Internet censorship circumvention. At its core, Infnote uses a decentralized storage system to store data on the network, similar to bitcoin.

\section{Blockchain}\label{sec:Blockchain}

Blockchain is a technology that is set to change how we currently conduct business in any given sector. Blockchain, in conventional terms, is a public ledger that records all events, transactions and exchanges that happen between parties or nodes in the network \cite{stanciu2017blockchain}. Bitcoin popularized the concept of blockchain, but blockchain as a baseline platform has far greater implications than bitcoin itself.


Data on a blockchain is stored on a chain of blocks, which is then accessed by users. It is guaranteed that the written information cannot be tampered with, since it relies on digital signatures and the hashing function. Unless the entire network fails or the cryptographic function is attacked, the information on the blockchain is secure and tamper-proof.


\subsection{Consensus Mechanisms}
In a blockchain system, the underlying assumption is that there is no centralized node and nodes generally do not trust each other. A consensus mechanism is a fault-tolerant mechanism to achieve necessary agreement on a single state over the network. In this section, we provide an introduction to some popular consensus mechanisms.

\paragraph{Proof of Work (POW)}
POW requires solving mathematical puzzles with easily verifiable answers. Bitcoin uses POW to achieve consensus\cite{nakamoto2008bitcoin}. The node that wishes to insert a block into the chain is called a miner. The mining process is based on a cryptographic hash function, which involves scanning for a value that, when hashed, results in a hash value that begins with a number of zero bits. Other nodes can easily verify the answer by hashing a single value. After a miner produces a satisfying hash value, they have the permission to insert a block (with transactions) into the chain. The bitcoin mining process currently needs a huge amount of computational resources as well as electricity to power the computers. The use of an application-specific integrated circuit (ASIC) can solve the mathematical puzzles much faster than a CPU and GPU, in both speed and efficiency, making it almost impossible for personal computers to join the mining process. In order to resist ASIC, new puzzles have been proposed, which not only rely on computational power, but also other computational resources, such as memory and disk space. In POW, nodes having sufficient computational resources are more likely to solve the mathematical puzzles and therefore have a higher chance of inserting blocks into the blockchain.

\paragraph{Proof of Stake (POS)}
POS states that a node needs to stake an amount of its tokens so it has the chance to insert blocks into the chain \cite{king2012ppcoin}. The more tokens a node stakes, the higher the chance of inserting blocks into the chain, because it is believed that the more tokens a user has, the less likely he will attack the network \cite{bentov2016cryptocurrencies}.

Instead of competing using computational resources, in proof of stake, nodes compete based on the number of tokens they stake, thereby reducing the energy requirements. Similar to POW, a node with tokens (rather than computational resources) has a higher chance of inserting blocks into the chain.

\paragraph{Delegated Proof of Stake (DPOS)}
Unlike POS, in which every node has the chance to insert blocks into the chain, DPOS only allows delegated nodes to insert blocks \cite{larimer2014delegated}. Delegated nodes are chosen by voting processes. Votes are weighted according to the number of tokens each voter stakes. The first tier of nodes (usually less than 100) which receive most of the votes will earn the right to insert blocks into the chain.


\paragraph{Proof of Authority (POA)}
In a POA network, only approved nodes can validate blocks and insert them into the chain \cite{Parity}. Unlike delegated nodes in the DPOS mechanism, approved nodes are not chosen by voting.

Currently, POA is mainly used in private networks, where every node knows the others and therefore trust approved nodes to maintain the chain. However, approved nodes have to maintain an uncompromised state given the power vested in them. Approved nodes need to gain their reputation through their work in the network. Any negative activity recorded can destroy the reputation of the approved node.

\paragraph{Practical Byzantine Fault Tolerance (PBFT)}
Numerous protocols have been proposed to solve the problem of Byzantine Fault Tolerance (BFT)\cite{lamport1982byzantine}. PBFT \cite{castro1999practical} is one solution, which can handle up to 1/3 of the malicious nodes. 

A block will be generated in a round, and each round can be divided into three phases: pre-prepared, prepared and committed. Each node has to receive confirmations from 2/3 of all nodes in order to enter the next phase \cite{castro1999practical}. Therefore, PBFT requires every node to be known to the network.

\paragraph{Delegated Byzantine Fault Tolerance (DBFT)}
DBFT is another solution to the BFT problem. The whole process is similar to PBFT, except only a small number of delegated nodes are voted to insert blocks into the chain \cite{NEO}.

\subsection{Consensus Mechanism Comparison}
Different consensus mechanisms have different advantages and disadvantages. We use the criteria given by \cite{vukolic2015quest} and \textbf{Table \ref{tab:consensus}} gives a comparison between them.

\begin{table*}
\centering  
\begin{tabular}{lcccccc}  
\hline
 &POW & POS & DPOS & POA & PBFT & DBFT\\
\hline 
\texttt{Identity Management} & Permissionless & Permissionless & Permissionless & Permissioned & Permissioned & Permissionless \\
\texttt{Latency} & High & High & Low & Low & Low & Low\\
\texttt{Throughput} & Limited & Limited & Excellent & Excellent & Excellent & Excellent\\
\texttt{Energy Saving}& No & Yes & Yes & Yes & Yes & Yes\\
\texttt{Scalability}& Excellent & Excellent & Limited & Limited & Limited & Limited\\
\texttt{Voting Process}& No & No & Yes & No/Yes & Yes & Yes\\
\hline
\end{tabular}
\newline
\caption{Consensus Mechanism Comparison}
\label{tab:consensus} 
\end{table*}

\paragraph{Identity Management (Rating: Permissionless or Permissioned)}
In POW, POS, DPOS and DBFT, everyone can download the code and participate in the network, generating new blocks by only knowing a single peer in the network. In POA and PBFT, only certain identifiable nodes can generate new blocks and each node needs to know the whole node list participating in the consensus.

\paragraph{Latency (Rating: Low or High)}
Latency is the amount of time for a transaction to be confirmed and accepted in the network. The blockchain systems based on POW need multi-block confirmations, causing high latency \cite{vukolic2015quest}. Current implementations of POS to either hybridize with POW or need checkpoints signed under the developer's private key, causing high latency. In DPOS, POA, PBFT and DBFT, the number of nodes participating in the consensus is small, leading to practical network-speed latencies.

\paragraph{Throughput (Rating: Limited or Excellent)} 
Due to the possibility of chain forks, POW has limited throughput \cite{vukolic2015quest}. Some of the implementations and variations based on POS outperform bitcoin when it comes to throughput but POS still has its limitations. EOS is a blockchain based on the DPOS consensus mechanism, which can support millions of transactions per second \cite{bach2018comparative}. PBFT and DBFT can sustain tens of thousands of transactions \cite{vukolic2015quest}. As the throughput of POA is bounded by hardware, not consensus, it has excellent throughput.

\paragraph{Energy Saving (Rating: Yes or No)}
Among all the consensus mechanisms, only POW needs an extensive amount of energy. Estimated annual electricity consumption for the entire bitcoin network currently is 73.12 TWh, about 30\% of the annual electricity consumption of all of Australia, as of Oct 2018 \cite{Digiconomist}.

\paragraph{Scalability (Rating: Limited or Excellent)} Here we examine the scalability in number of nodes participating in the consensus mechanism. POW and POS have excellent scalability, easily supporting thousands of nodes, while PBFT has only been tested on a small number of nodes \cite{vukolic2015quest}. The DPOS, POA and DBFT mechanisms rely on a few delegated or approved nodes.

\paragraph{Voting Process (Rating: Yes and No)}
Here we examine whether the nodes need to vote in order to achieve a consensus about writing or inserting blocks. In DPOS, PBFT and DBFT, the nodes participating in the consensus vote for a block, deciding on whether to insert it into the blockchain.

\subsection{Incentive of Blockchain System}
Most public blockchain systems rely on cryptocurrency to motivate their participants. The cryptocurrency can be exchanged into fiat money through exchange platforms. The blockchain system maintains a transaction ledger, where the balance of each account can be calculated. The cryptocurrency can be transfered to another account through a transaction, which will be written into the ledger. By generating new blocks, miners can receive transaction fees and block rewards. It is the cryptocurrency system which keeps most public blockchain systems working well, since any misbehavior would cause the loss of cryptocurrency and therefore fiat money. 

\section{Design And Implementation}\label{sec:Designandimplementation}

Infnote is a general information sharing platform based on blockchain that can support various applications, such as a portal to share information, a blog to write articles, or a forum to discuss topics. The features that distinguish Infnote from existing platforms is that we are able to provide users access and publish contents in censorship-driven countries without costing any cryptocurrency, preserve all history of content, verify and trust that the source of the content is from the original author, provide assurance that data will not be tampered with or lost once published and offer access to the platform regardless of what type of device the user owns. This is made possible through the use of P2P, blockchain and our unique architecture. In this section, we discuss the design of Infnote from consensus mechanism choices, a platform designed from the ground up, protocol usage, architecture design and the overall technology it relies on.

\subsection{Infnote Chain}

It is possible to directly use or fork popular blockchain systems such as bitcoin or Ethereum and build Infnote upon them. However, in order to transact in these blockchain systems, crytocurrency is a necessity, which is a huge barrier to entry for many users. The open letter sent to Ethereum against Chinese government censorship costed \$0.52 worth of cryptocurrency \cite{MetooBlockchain}. Keeping this barrier in mind, we decided against any monetary element. We assume that, as a free-for-all information sharing platform against censorship, freedom of speech is an already strong enough incentive for people to join and contribute to Infnote. We think that this assumption could be true, similar to Wikipedia, which does not explicitly credit authors for their work\cite{forte2005people}.

It is also possible to modify the existing private blockchain systems without any monetary element to build Infnote. However, in order to gain more flexibility, we decide to design and implement Infnote from the ground up.

\begin{figure*}
\centering
\begin{minipage}[t]{0.49\textwidth}
\centering
\includegraphics{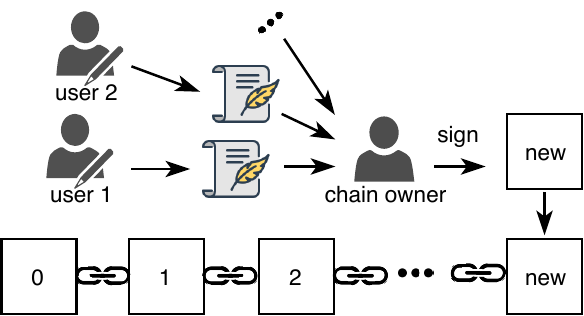}
\caption{The process of publishing a post and inserting it into the blockchain.}
\label{fig:publishtoblockchain}
\end{minipage}
\begin{minipage}[t]{0.49\textwidth}
\centering
\includegraphics{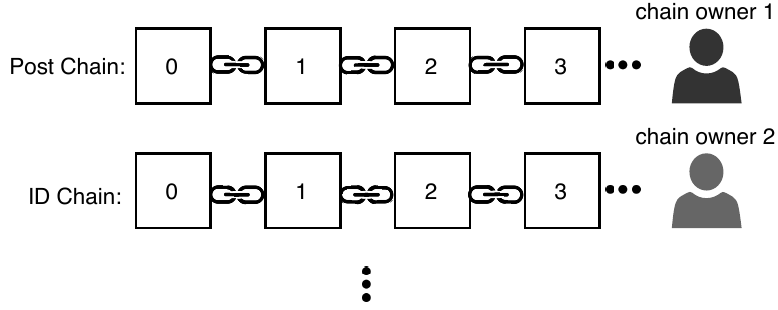}
\caption{Multi-chain structure}
\label{fig:multichain}
\end{minipage}
\end{figure*}

Infnote utilizes blockchain technology to store information. When a user wishes to publish a post on the platform, the post will be signed with the user's private key. Later, the post will be bundled with other posts and additional information (like timestamp etc.) together into a block. The \textit{chain owner}, who has the authority to insert blocks into the blockchain, will sign the block with his private key. \textbf{Figure \ref{fig:publishtoblockchain}} demonstrates the process of posts inserted into the blockchain. The block will be broadcasted to the P2P network, and every node in the network will verify it. We want to emphasize that, although a chain owner has the power to decide whether to insert a block, the chain owner's power is limited, because  no one will be able to remove it after insertion, due to the nature of blockchain.

\paragraph{Cryptography}
We utilize a digital signature scheme implemented using \textit{ECDSA} with \textit{secp256k1} curve \cite{johnson2001elliptic} and a cryptographic hash function \textit{SHA-256} \cite{pub2012secure}, the same building blocks as bitcoin.

\paragraph{Block}
A typical block structure contains the following attributes: ChainID, Height, Time, PrevHash, Hash, Signature and Payload. ChainID is the hash value of the chain owner's public key. Height is the current block height. Time is the generating time of the current block. PrevHash is the hash value of the previous block while Hash is the hash value of the current block. Signature is the block signature signed by the chain owner's private key. An applications' data can be put into Payload. A chain owner will generate a new block signed with his or her private key and broadcast it to the nodes that are connected to it. Then, every node which receives the block will verify it by checking the correctness of ChainID, Height (to make sure there are no two blocks with same height in the same chain), Time (time must be later than the generating time of the previous block), PreHash, Hash and Signature.

\paragraph{Multi-chain}

Infnote uses a multi-chains architecture, which means there are several independent parallel chains. Each chain is controlled by its chain owner. Multi-chains architecture can be used in various independent applications or different functions in an application. \textbf{Figure \ref{fig:multichain}} demonstrates an example of a simple discussion forum, in which posted data is stored in the Post Chain while user data is saved into the ID Chain which can be shared among other applications.

Everyone can become a chain owner simply by creating a new chain. However, whether it will be maintained by enough nodes depends on the reputation of the chain owner and the quality of the information in the chain. The community of Infnote would maintain a default list of chains recommending the users to follow. With this mechanism, the chain owners are given the incentive to follow the code of conduct. Any chain owner who violates the general rules set by the community would be removed from the default list. If the user disagrees with the community's decision, a user can simply override the default list to maintain or drop certain chains. In this model, each participant only has limited power and the ultimate decision is made by the users themselves.

One type of chain is an ID chain that is used to store all the users' information, which can be shared among different applications. By utilizing an ID chain, a user can map its public key to an easy-to-remember unique user name. Users can also store additional personal information on the ID chain. Note that Infnote does not restrict the number of ID chains or the owner of the ID chain. Different ID chains would compete with each other and only the chains which gain users support would survive. 

\paragraph{Consensus mechanism}

Infnote, as an information sharing platform, must ensure its quality of content, such as not allowing machines or bots to automatically send advertisements onto the platform. As Infnote does not include any currency system, sending advertisements to the platform is almost free. The two common consensus mechanisms POW and POS are not compatible with Infnote, since there is no guarantee of which node will generate the next block and insert it into the blockchain, and therefore no guarantee of what kind of posts will be published on the platform. PBFT does not allow many nodes to participate in the network\cite{vukolic2015quest}, and thus does not suit Infnote's requirements either. In DPOS and DBFT, only delegated nodes can insert blocks into the blockchain. However, when determining whether a post should be published on the platform or not, delegated nodes may have conflicts, making it harder to reach a consensus. This would cause a significant delay in writing information into the blockchain. Therefore, Infnote uses POA as its underlying consensus mechanism. POA only allows authorized nodes to insert blocks into the blockchain. Only a chain owner can insert blocks into a chain and therefore control the information on the platform.  

Just like a miner in bitcoin, a chain owner's role is to generate new blocks signed with his private key and broadcast it to the nodes which are connected to him. Due to the append-only property of blockchain, the chain owner's power is limited. Once the chain owner decides to insert a block into the blockchain, it will be broadcasted to the P2P network, and thus it becomes impossible to remove from the blockchain. The chain owner can still soft delete a post by inserting another block to mark the deletion of that post, but the history will be permanently recorded in the blockchain and there is no way to remove it.

\paragraph{Fork}
It is possible that a chain owner signs two blocks, causing the blockchain to diverge into two paths, like a fork in bitcoin \cite{lin2017survey}. However, this is strictly prohibited in Infnote. If any node detects that two blocks of the same height are signed with the correct signature of the chain owner, it will stop trusting the chain owner and stop broadcasting his or her blocks. Without the support of the P2P network, the chain owner cannot send the information out. In a scenario where the chain owner's private key is stolen, this provides a termination method for the chain owner to permanently close his or her blockchain.

\paragraph{Implicit reputation system}
Unlike traditional information sharing platforms, like Facebook or Twitter, where the identity of the owners is open to all. In Infnote, a chain owner can choose to hide his or her identity by using an anonymous communication technology like Tor \cite{syverson2004tor}. Each chain owner gains his or her reputation on the network by the work they conducted so far. Even if a chain owner decides to hide his or her identity, users are able to observe the chain owner's behavior in the blockchain and decide whether or not to use his or her services. Generally, it is expected that the higher the reputation, the more peers will join the network.

\subsection{Nodes}
We fully expect multiple types of devices to join the network, such as laptops, desktops, servers, smart-phones and so on. The front end interface can be through a web browser, a stand-alone program, or a smart-phone app. However, different devices have different capabilities, so it is necessary to analyze the features of each kind of device and design different strategies for them. In Infnote, there are two kinds of nodes representing its devices, full nodes and light nodes. 

\paragraph{Full Nodes}
Personal computers and servers can be full nodes. As with bitcoin, full nodes are devices that have sufficient bandwidth and computational resources to support all the functions of Infnote, which include storing all the data in the blockchain, providing logic to view and publish the content, acting as a server by listening for connections, and providing services to clients. People and organizations can run full nodes by using their spare resources. A full node client can be run on a desktop, a server, or a virtual machine. Full node client is implemented using Go, which provides us cross-platform interoperability. The database layer is powered by SQLite.

\paragraph{Light Nodes}
Many devices, such as smart-phones or web browsers cannot be full nodes, due to limited resources and processing power. Hence, they must rely on the full nodes to provide comprehensive services. At the same time, light nodes can still use their limited resources to contribute to the system.

Smart-phones usually do not have much storage space; therefore it is unreasonable for a smart-phone to store all the data in the blockchain. It is also unlikely that a smart-phone will run the Infnote software for a long time. Most of the smart-phone platforms, however, allow software to make an Internet connection, making it possible for a smart-phone to join the P2P network. A smart-phone device can cache some recent blocks and broadcast them to the P2P network. The phone user is able to view the data stored in the recent blocks. We have developed an iOS app implemented using Swift to demonstrate the functionalities and features of a light node.

Web browsers are restricted environments. A program written in JavaScript can be run on web browsers, but with more restrictions than a phone app. We had to resolve two main issues: storage and the communication protocol. Before HTML5, application data had to be stored in cookies, which would be sent to the server upon on every request. Web storage is a more secure method for storing data locally, supporting larger amounts of data, while the data will never be sent to the server. Today, most web browsers support web storage. As with smart-phones, it is impossible for web browsers to store the entire blockchain data, but by utilizing web storage, the program running on web browsers can cache some recent blocks and the user can view the information in those blocks. The communication protocol is strictly restricted in web browsers. Since UDP and TCP protocols are not directly allowed in most web browsers, Infnote uses Websocket as its communication protocol, which is currently supported by most major browsers. The current implementation of Infnote's web client can run on any web browser without additional software.

\paragraph{Multi-layer Structure}

\begin{figure}
\centering
\includegraphics{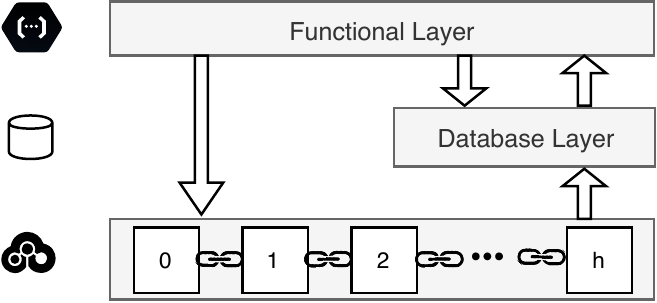}
\caption{Three layer structure}
\label{fig:threelayers}
\end{figure}

Infnote has implemented a number of user-friendly functions on top of the blockchain, which are supported by a multi-layer structure. A full node has three layers, while a light node may only have two layers. \textbf{Figure \ref{fig:threelayers}} shows a typical three-layer structure. An arrow in the figure represents the direction of a data flow. The three different layers are described below.

\begin{enumerate}
\item
Blockchain Layer: All nodes have this layer. The blockchain layer serves two purposes: it stores all the data of Infnote in sequence and provides the advantages of using blockchain. A full node is expected to store all the blocks, while a light node is only expected to cache a few recent blocks, due to limited storage space.

\item
Database Layer: All full nodes have this layer. It is necessary to reorganize the data into the database, since relying only on sequence data, a full node cannot provide services efficiently. Light nodes may also run a database to improve the efficiency. 

\item
Functional Layer: Both types of nodes have this layer. This layer provides various operations to users, like publishing or viewing an article in Infnote. The function layer verifies the operation based on the data in the database, but updates are applied to both the database and the blockchain accordingly. Since a light node is only expected to cache some recent blocks, the functions a light node can provide are limited.
\end{enumerate}

\subsection{Network}
A network allows nodes to interact with each other. In this section, we discuss the specific characteristics of our network.

\paragraph{Decentralized Network}
The P2P network plays a vital role when developing the entire system and laying down the architecture. Similar to bitcoin, Infnote's architecture does not rely on a centralized server. For a censorship-resistant platform, this is a necessary condition, since any single server would easily be blocked by censors.

\paragraph{Broadcasting Blocks}
Similar to bitcoin, whenever chain owners generate a block or nodes receive a new block, they will immediately send the new block out to the peers which they have direct connections with so that every node can obtain the new block in a short time. This follows the same principle as the publish-subscribe pattern, where publishers (chain owners) send blocks and subscribers receive these blocks. This feature allows nodes to automatically obtain new posts in Infnote in a short time.

\paragraph{Peer Discovery}
Peer discovery is extremely crucial for a P2P network to circumvent Internet censorship. How to find the initial peer when a new node wants to participate in the network is known to be a difficult task. For a pure decentralized P2P network, the only way that is guaranteed to succeed is to search on the Internet and send a handshake message to millions of addresses hoping to find one peer who has already joined the P2P network. However, this is not realistic. A more practical solution is to centralize, making the initial peer discovery the weakest link in the entire system. Authorities may simply block the initial seeds and thus prevent new nodes from joining the network. Infnote provides several methods for a node to initially find peers in the network to relieve this issue.

\begin{itemize}

\item Hard-coded nodes: The developing community can hard code several recommended nodes in different geographies. This method, however, may increase the workload of those nodes and they are likely to be blocked. 

\item DNS Seeding: DNS seeding servers run a web crawler exploring the stable nodes in the P2P network and maintain a list of them. Whenever a node request is sent to a DNS server, it will return multiple node addresses. The DNS protocol is a light protocol; therefore, this will not result in a heavy workload for DNS seeding servers. However, the DNS seeding servers might also be blocked.

\item From other nodes: Once a node joins the P2P network, the node can send requests to other nodes asking for more nodes' addresses. 

\item Address database: A node can store the addresses of nodes in its local database. On the next runtime, the node may not need to do the initial peer discovery given that nodes in the address database are still available.

\item User-specified address: The users can manually specify a node address in the software. The users can enter the address or simply scan a QR code. Although this method seems less efficient, it is the hardest for authorities to stop initial peer discovery. This method allows users to join the P2P network by relying on real-life connections, which seems like the only solution in countries with pervasive censorship .

\end{itemize}

\paragraph{Obfuscation}
Censors may use DPI to detect the protocol deeper inside the network packets. By using obfuscation, our goal is to avoid detection of Infnote packets. Infnote currently uses three approaches:

\begin{itemize}

\item Random Port: The regulators may ban some ports, so that certain services will not work. For example, not allowing packets through port 80 can prevent access to HTTP websites. Infnote can support the use of random ports to communicate between each node, so that regulators inspecting and blocking port numbers will not be able to block Infnote.

\item Mimicry: With this method, packet payloads are made to look like something that would be allowed by the DPI. A common example would be making the payloads look like HTTP packets, which are rarely blocked, because of its ubiquity \cite{dixon2016network}. Infnote directly uses WebSocket as its underlying communication protocol. Similar to HTTP, WebSocket is a commonly used protocol and, therefore should not be blocked by censors.

\item Encryption: Similar to HTTPS, the WebSocket protocol supports encrypted connections, indicated by the prefix wss in the URI. By using encrypted connection, the censor would not be able to obtain the content of the packets by intercepting network traffic.

\end{itemize}

\paragraph{Pseudo-anonymity}
Similar to bitcoin, Infnote provides users with pseudo-anonymity. In countries where substantial censorship rules exist, a user's identity may need to remain anonymous. If more nodes join the P2P network, the difficulty of finding the owner of a post would increase.

For users who need a higher level of anonymity, they can combine the Onion Router (Tor) \cite{TorProject} with Infnote. Once a user uses Tor to make a connection, the data packets will be relayed multiple times over distinct intermediary servers and each server only knows limited information about the packets, making it extremely difficult to trace back to the source. 

\subsection{Modes}

Infnote, depending on needs and requirements, can work in different modes. In essence, Infnote provides a solution for different degrees of Internet censorship. 

\begin{figure*}
\centering
\begin{minipage}[t]{0.49\textwidth}
\centering
\includegraphics{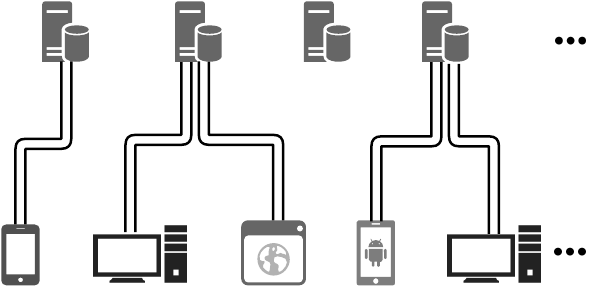}
\caption{Direct Connect Mode with multiple servers}
\label{fig:DirectConnection}
\end{minipage}
\begin{minipage}[t]{0.49\textwidth}
\centering
\includegraphics{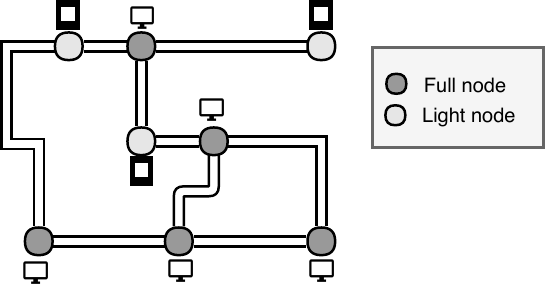}
\caption{Peer-to-Peer (P2P) Mode}
\label{fig:P2Pmode}
\end{minipage}
\end{figure*}

\paragraph{Direct Connect Mode}
This mode is the same as the traditional client-server architecture, in which the client is the requester while the server is the service provider. In a client-server model, the server will handle requests and return the information to the client \cite{kambalyal20103}. In Infnote, a full node could be a server, which can provide comprehensive functions. In an area where the server can be directly accessed by users, direct connect mode is the most efficient method. The server normally has a powerful computing capacity and more network bandwidth, and can therefore support more clients and provide more functions. The user can easily connect to the server by using HTTP or HTTPS protocol. By using Tor, the user can even establish anonymous communication with the servers \cite{TorProject}.

Owing to all the data being stored in the blockchain, the servers are able to provide comprehensive services based on the data in the blockchain. Any node which has adequate capacity can download the blockchain and become a server to handle requests from the client. This feature enables Infnote's architecture to support multiple servers, making the system much more robust and censorship resistant. \textbf{Figure \ref{fig:DirectConnection}} demonstrates an example of multiple server handling requests from multiple clients.

\paragraph{Peer-to-Peer (P2P) Mode}
In areas with substantial censorship or above levels, a direct connection may not work. The common approach would be to use a proxy service. However, if the whitelist type of Internet censorship is being implemented, the proxy server may not be allowed access. 

Infnote can utilize the P2P network to transfer or receive data. Every node in the P2P network would actively broadcast and receive blocks. Once a node receives a block, the node is able to extract and validate the data in the block. Similar to bitcoin, in which the user can send a transaction to any full node to then be broadcasted to the whole network and written into the blockchain, the user can send their data into the P2P network to be permanently written into the blockchain later. \textbf{Figure \ref{fig:P2Pmode}} demonstrates a possible scenario of network structure in P2P mode.

\paragraph{Without Internet Mode}

In extreme situations, access to the Internet may be disconnected partially or fully due to political reasons, similar to what happened in Egypt during the Arab Spring. Although it is impossible for citizens to access the Internet, the infrastructure of the Internet can still be utilized. Every router can establish an internal network, so that as long as nodes are in the same internal network, they can still send blocks to other nodes. Each smart-phone can be a data truck that transfers the blocks in different internal networks.


\subsection{Preventing Attacks}
In order to prevent messages flooding into the system, chain owners could use all techniques that have been implemented in traditional websites, such as utilizing CAPTCHA or binding with users' social media accounts. Additionally, Infnote uses the same set of cryptography building blocks like bitcoin, making it possible to verify users' accounts and even requiring payments in bitcoin blockchain. For the malicious chain owners or adversaries who have made up identities to be chain owners, a node can simply disconnect itself from the P2P network automatically, based on local settings. As there is no restriction of becoming a chain owner by creating a new chain, all nodes will not automatically maintain new chains, unless users or the developing community decide to maintain them, preventing useless or low quality chains.

\section{Evaluation}\label{sec:Evaluation}

\subsection{Throughput and Latency}
We evaluate our system quantitatively based on two aspects of the system: throughput and latency. Throughput is how many posts the system can handle per second. Latency is the amount of time it takes for a block to be confirmed by all nodes on the P2P network. For testing, we assume that the size of a post in Infnote is 250 bytes. We also assume that the size of a block is 1 megabyte. Due to the architecture being different to all other projects mentioned in section \ref{sec:RelatedProjects}, it is difficult to compare with them directly. Even so, we will reference some results of bitcoin to show that our implementation is feasible.

Throughput: Our experiment shows that Infnote's throughput can reach approximately 150,000 posts per second, running the Go version of the Infnote program on a 64-bit machine with an Intel Core i7 CPU @2.70GHz with 16GB RAM. The result could be further improved by deploying better hardware or optimizing the code. Bitcoin, which takes approximately ten minutes to confirm a block, achieves only 7 transactions per second maximum throughput \cite{croman2016scaling}.

Latency: We simulated a global environment using nodes that were spread across the world geographically. In a test environment, as it is impossible to deploy a P2P network on a large scale due to limited resources, we speculate on the performance of such a system by using only a small number of nodes. We utilize ten nodes in different geographic regions around the world, with eight full nodes and two light nodes. \footnote{The nodes are located in Tokyo (node No.1), Singapore (node No.2), Kuala Lumpur (node No.3), Sydney (node No.4), Mumbai (node No.5), Hong Kong (node No.6), Dubai (node No.7), Hong Kong (node No.8), Silicon Valley (node No.9) and Hong Kong (node No.10).} Node No.8 is a light node running on an iPhone 8, and node No.10 is another light node running on a JavaScript program of Infnote on a Chrome web browser. Node No.1 is the chain owner who generates new blocks. We tested our system in both a star topology network and a linear topology network. 

\begin{figure}
\subfigure[latency with different number of nodes]{
\label{fig:latency1}
\includegraphics[width=0.22\textwidth]{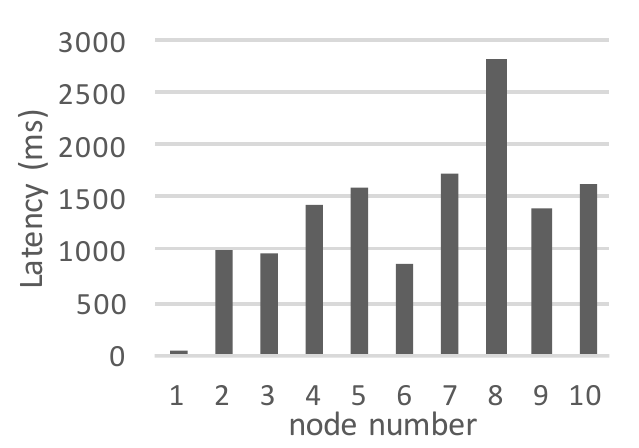}}
\hfill
\subfigure[latency with different network diameters]{
\label{fig:latency2}
\includegraphics[width=0.22\textwidth]{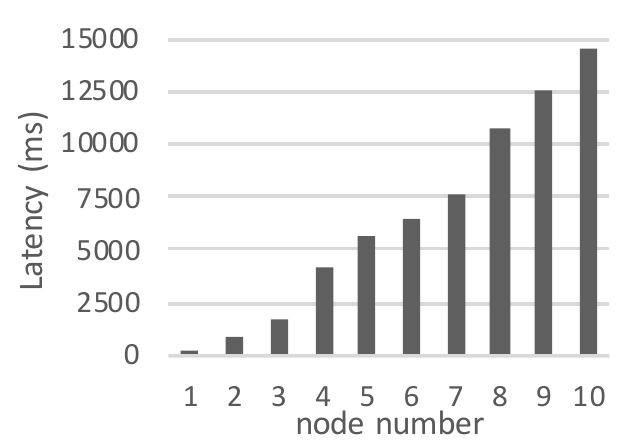}}
\caption{Latency of the Infnote}
\label{Fig.main}
\end{figure}

\textbf{Figure \ref{fig:latency1}} shows the results in the star topology network. On average, the latency is 1.3 seconds, for every node in the network to receive a 1 megabyte block, which basically matches the network latency. \textbf{Figure \ref{fig:latency2}} shows the results in the linear topology network. The experiment aims to represent the results of a large network with different network diameters. The first node in the chain is the chain owner. Infnote takes 14.6 seconds to transmit a 1 megabyte block in the entire network with 10 diameters. For the bitcoin network at that time with around 3500 reachable nodes, one result shows that the median latency is 6.5 seconds whereas the mean is at 12.6 seconds and after 40 seconds there are still 5\% of nodes that have not yet received the block \cite{decker2013information}.

\subsection{Effectiveness}
\label{sec:Effectiveness}
In this part, we analyze the effectiveness of Infnote on different levels of internet censorship. We assume that, except on level 5, a user will be able to establish Infnote connections via the Internet among one another within a censorship region. Although a censor can deploy advanced censorship methods like DPI to scan the detailed content of packets, a user can encrypt and obfuscate packets to look like something that would be allowed by the DPI, such as HTTP or HTTPS packets, which are rarely blocked \cite{dixon2016network}. 

\paragraph{Level 1 and Level 2}
In a little or selective censorship environment, it is not necessary to use Infnote as a circumvention tool. Infnote is a way of preserving the history of all users and site owners permanently and providing everyone the ability to fork its database, with acceptable overheads. 

\paragraph{Level 3}
In a regime with substantial censorship, since Infnote does not rely on a single point, it is extremely hard to completely shut down all the Infnote nodes, providing there are sufficient nodes. We want to emphasize that the chain owner could use his private key to sign a block and submit it to any node in the Infnote network without relying on a fixed server, which is similar to a bitcoin user to send a bitcoin transaction without relying on a fixed device. All user and chain owners can combine Infnote with Tor to achieve higher level of anonymity. Due to a whitelist not having yet been deployed by the firewall on this level, the chain owner could use servers located in a none or less censorship environment to mitigate the risks of being traced.

\paragraph{Level 4}
At the pervasive censorship level, since a whitelist has been deployed by the firewall, it is impossible to obtain sensitive information from outside of censored regions. However, Infnote could provide a solution to build a P2P information sharing network inside the censored region. On this level, a chain owner has to be located in the censored region and a single chain owner could be traced and isolated, but everyone can easily become a chain owner by simply creating a new chain.

\paragraph{Level 5}
In extreme situations, access to the Internet may be disconnected. Although it is impossible for citizens to access the Internet, the infrastructure of the Internet can still be utilized. Every router can establish an internal network, so that as long as nodes are in the same internal network, they can still join the P2P network and send blocks to other nodes. Each smart-phone can be a data truck that transfers the blocks via different internal networks.

\section{Demonstration}
\label{sec:Demonstration}

We demonstrate a simple discussion forum based on Infnote \footnote{Demo video: https://www.youtube.com/watch?v=027QOEJRqKY}. Currently, it supports two platforms: web browser and iOS. \textbf{Figure \ref{fig:infnote_io}} and \textbf{Figure \ref{fig:iOS_bbs}} present the interfaces of the discussion forum containing the basic functions as a discussion forum, such as viewing and posting an article, registering a new user and logging onto an existing account. Similarly to bitcoin, a user needs to use his or her private key to log into the system. The app on an iOS platform provides more functions for the users, such as logging into an account by scanning a QR code that contains the private key and saving the private key to the iCloud service.

When the iOS app is running, it automatically becomes a light node that is able to receive and broadcast blocks in real time. \textbf{Figure \ref{fig:iOS_block}} shows the details of a block which has been saved to the local database of the phone. It is also possible to receive and broadcast the blocks by a light node client implemented by JavaScript running in a web browser environment, as displayed in \textbf{Figure \ref{fig:infnote_browser}}.

Software for full nodes is implemented using Go, which provide all necessary operations for a full node as a normal user or chain owner. This includes broadcasting and receiving new blocks, creating a new chain or a block, deleting a chain, querying a block in a chain and maintaining a new chain, in the command line environment. 

{%
\setlength{\fboxsep}{0pt}%
\setlength{\fboxrule}{0.1pt}%

\begin{figure*}
\centering
\begin{minipage}[t]{0.48\textwidth}
\centering
\fcolorbox{gray}{gray}{\includegraphics[width=0.95\textwidth]{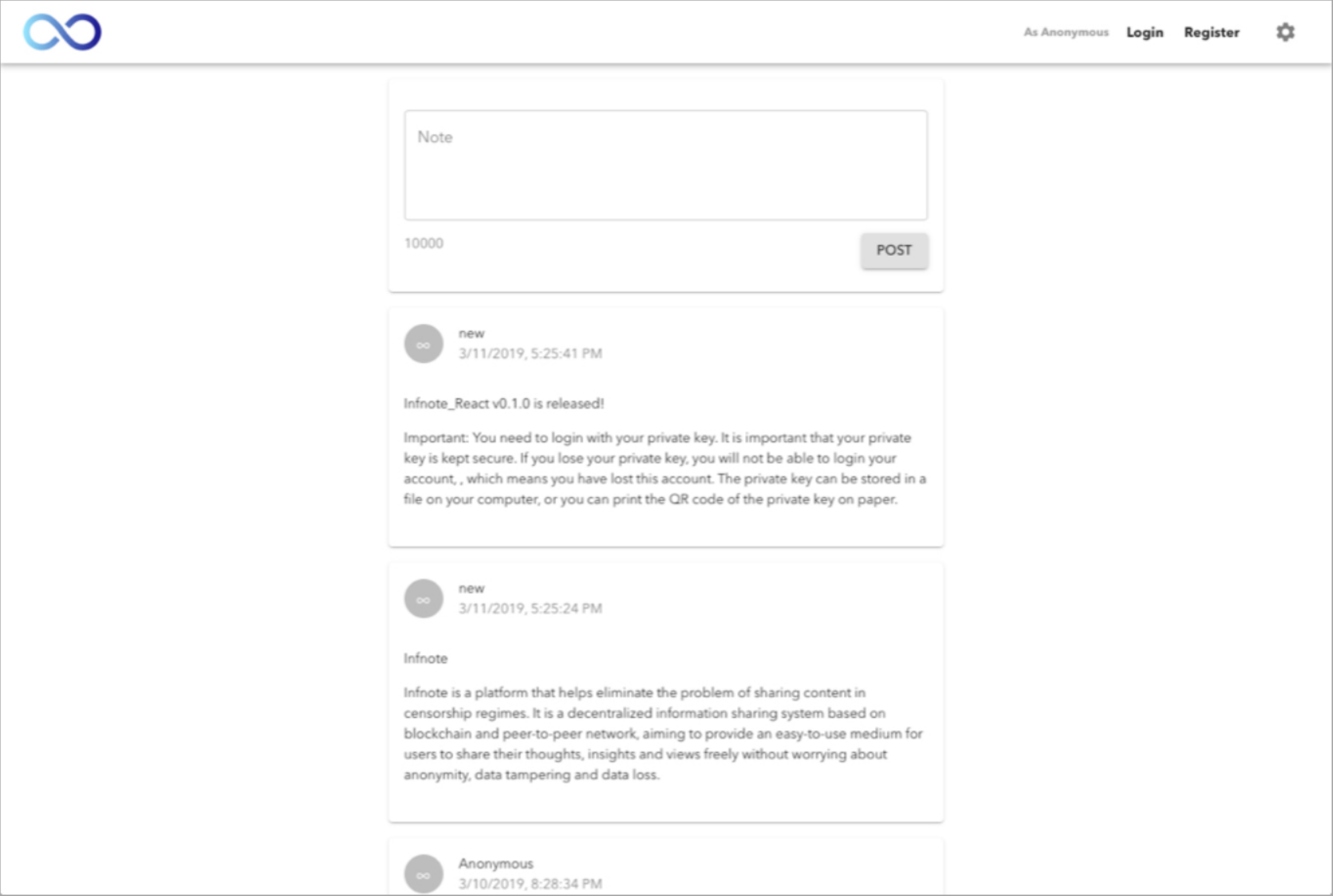}}
\caption{Web interface of a simple discussion forum based on Infnote}
\label{fig:infnote_io}
\end{minipage}
\begin{minipage}[t]{0.48\textwidth}
\centering
\fcolorbox{gray}{gray}{\includegraphics[width=0.95\textwidth]{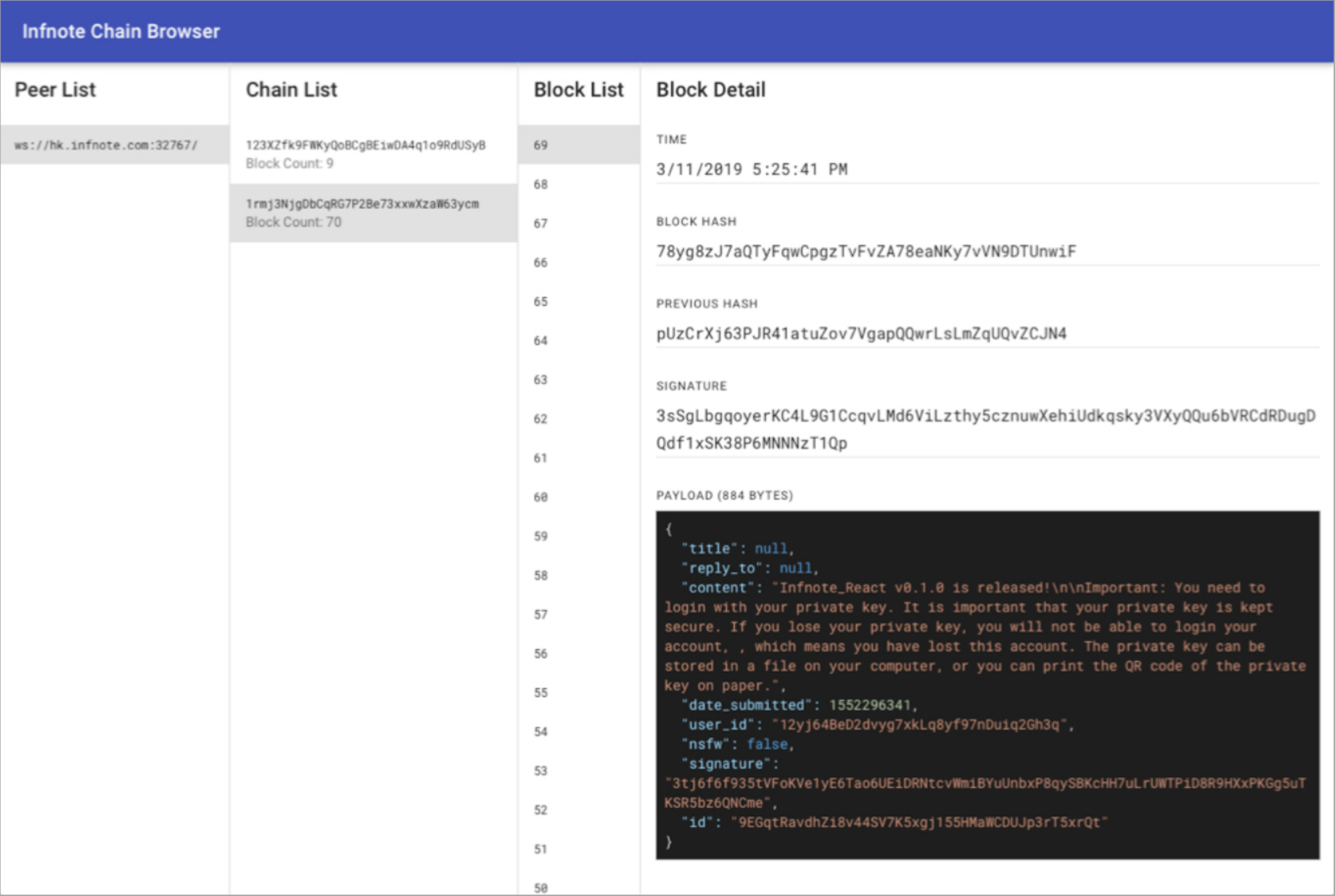}}
\caption{Infnote blockchain browser}
\label{fig:infnote_browser}
\end{minipage}
\end{figure*}

\begin{figure}
\subfigure[Discussion forum]{
\label{fig:iOS_bbs}
\fcolorbox{gray}{gray}{\includegraphics[width=0.20\textwidth]{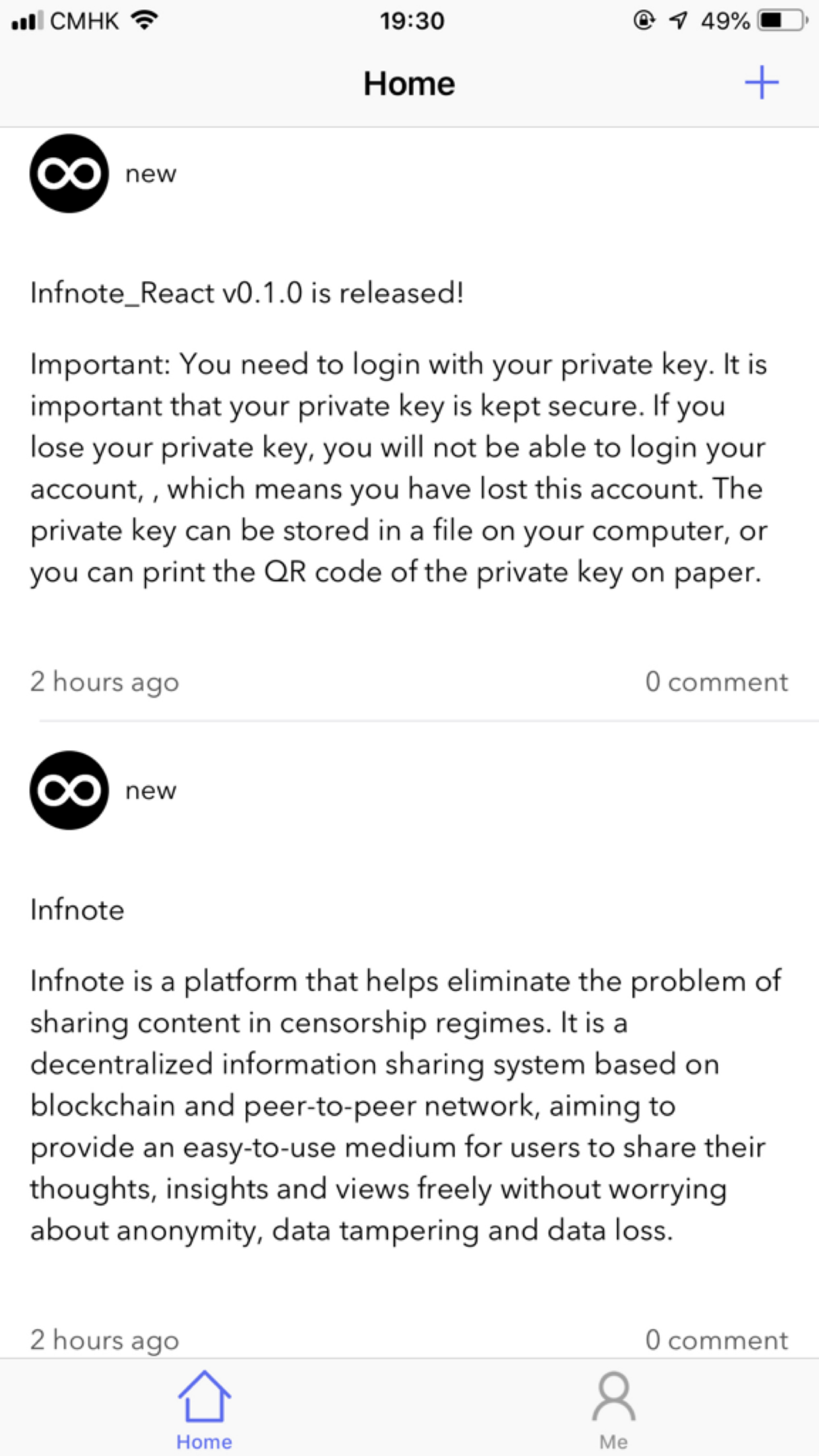}}}
\hfill
\subfigure[Blockchain browser]{
\label{fig:iOS_block}
\fcolorbox{gray}{gray}{\includegraphics[width=0.20\textwidth]{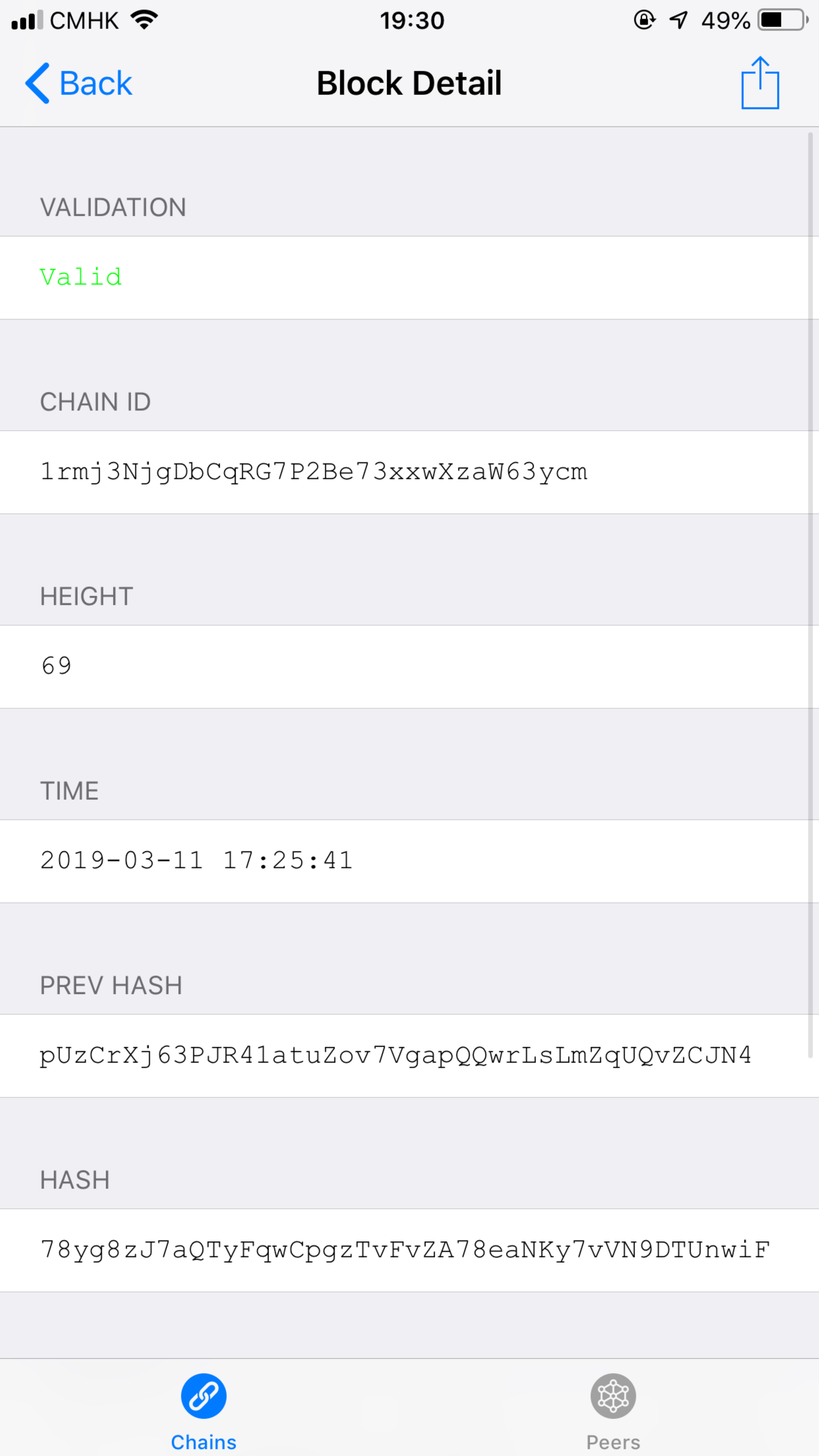}}}
\caption{Infnote App on iOS }
\end{figure}

}%

\section{Future Work}\label{sec:futureWork}

In the current implementation of Infnote, in direct connect mode, the servers (full nodes) can provide comprehensive services to clients (light nodes) and access to content that is on the blockchain. However, the servers can feed the clients with wrong information that may not have been written to the blockchain at all. In future work, we plan to update the architecture to use authenticated data structures (ADS) where responders need to also send back proof that the content came from the blockchain it claims \cite{miller2014authenticated}.

\section{Conclusion}
\label{sec-conclusion}

In this paper, we began by setting objectives to provide a platform to users around the world to share their views and opinions with the underlying assumption that the content shared will remain intact, unchanged and be protected.

We defined a few levels and types of censorship that can be used to put different countries into categories for comparison purposes and for future research. We also compared and contrasted existing circumvention technologies that bypass blockages and filters. Each technology was analyzed on the basis of its effectiveness to bypass all levels of censorship. 

Blockchain, as an underlying technology, met several of the objectives. After careful analysis and comparison, we came to the conclusion that POA works with Infnote's plans and vision. To store posts into the blockchain, a multi-chain architecture comprising of full nodes and light nodes is developed. 

Infnote program with its unique construction allows it to create a platform for its users which is decentralized, tamper-proof, safe and open to everyone. From the proof of concept of Infnote, the evaluation stage showed immense promise. The Infnote program achieved significant throughput (posts per second) and low latency while spreading the blocks around the world. 

\bibliographystyle{IEEEtran}
\bibliography{ref}



\end{document}